\documentclass[namedreferences]{SolarPhysics}
\usepackage{graphicx}
\usepackage{spr-sola-addons}
\usepackage{epstopdf}

\begin{document}
\begin{opening}
\title{Cross-calibrating Sunspot Magnetic Field Strength Measurements from the McMath-Pierce Solar Telescope and the Dunn Solar Telescope}

\author[addressref={aff1},corref,email={fwatson@nso.edu}]{Fraser T. Watson}
\author[addressref={aff2}]{Christian Beck}
\author[addressref={aff3}]{Matthew J. Penn}
\author[addressref={aff2}]{Alexandra Tritschler}
\author[addressref={aff1}]{Valent\'{i}n Martinez Pillet}
\author[addressref={aff3}]{William C. Livingston}

\address[id=aff1]{National Solar Observatory, Boulder, CO}
\address[id=aff2]{National Solar Observatory, Sunspot, NM}
\address[id=aff3]{National Solar Observatory, Tucson, AZ}

\runningtitle{Cross-calibrating McMP and DST sunspot data}
\runningauthor{Watson et al.}

\begin{abstract}
In this article we describe a recent effort to cross-calibrate data from an infrared detector at the McMath-Pierce Solar Telescope and the Facility InfraRed Spectropolarimeter (FIRS) at the Dunn Solar Telescope. A synoptic observation program at the McMath-Pierce has measured umbral magnetic field strengths since 1998 and this data set has recently been compared with umbral magnetic field observations from SOHO/MDI and SDO/HMI. To further improve on the data from McMath-Pierce, we compare the data with measurements taken at the Dunn Solar Telescope with far greater spectral resolution than has been possible with space instrumentation. To minimise potential disruption to the study, concurrent umbral measurements were made so that the relationship between the two datasets can be most accurately characterised. We find that there is a strong agreement between the umbral magnetic field strengths recorded by each instrument and we reduce the FIRS data in two different ways to successfully test this correlation further.
\end{abstract}
\keywords{Active Regions, Magnetic Fields; Magnetic Fields, Photosphere; Sunspots Magnetic Fields}
\end{opening}

\section{Introduction}
The synoptic gathering of solar data has become a major component of solar physics research in recent times as we strive for a better understanding of how the Sun evolves over timescales of a solar cycle and longer. The best example of this is the effort currently being made to standardise the sunspot number over its 400 year observation history \citep{2014SSRv..186...35C}. However, a more recent observation that has become a standard synoptic measurement is the magnetic field of various features on the Sun. The importance of this has been made clear by the constant development of world-class synoptic instrumentation to record field strengths, such as magnetograms from the {\it Michelson-Doppler Imager} (MDI; \citealp{1995SoPh..162..129S}) on the {\it Solar and Heliospheric Observatory} (SOHO) satellite, the {\it Helioseismic and Magnetic Imager} (HMI; \citealp{2012SoPh..275..207S}) on the {\it Solar Dynamics Observatory} (SDO) and the {\it Global Oscillation Network Group} (GONG; \citealp{1996Sci...272.1284H}) instruments. In addition, many authors have undertaken the task of comparing results from a number of instruments dedicated to measuring magnetic fields on the Sun. \citet{2008SoPh..250..279D} compared magnetic fields measured using different spectral lines at the same observatory, as well as using the same spectral line at different observatories. A more detailed study by \citet{2014SoPh..289..769R} involved a comparison of magnetic field measurements from seven different ground and space based instruments. This study found general consistencies but was not able to settle on a `ground truth'. Various techniques for measurement of photospheric magnetic fields are detailed in articles such as \citet{2012SoPh..276...43D} (multi spectral line spectropolarimetry) and \citet{2013A&A...552A..50M} (Milne Eddington inversion of spectral line observations). These studies dealt mainly with the determination of the magnetic field strength on the complete solar disk using visible spectral lines. Because of the smaller Zeeman splitting at visible than infrared wavelengths and the presence of unresolved magnetic structures, there are ambiguities in the determination of line-of-sight magnetic flux or magnetic field strength. Currently no synoptic full-disk observations at infrared wavelengths are available.

One of the few synoptic programs measuring solar magnetic fields at infrared wavelengths, explicitly those in sunspot umbrae, has taken place at the McMath-Pierce Telescope facilty on Kitt Peak, AZ since 1998. This series of measurements, using the instrument called BABO \citep{BABO}, currently covers sixteen years of sunspot observations and, most importantly, measures magnetic field strengths using a Zeeman splitting technique that returns the true strength at source. MDI and HMI produce magnetograms that give the line-of-sight magnetic field strength, introducing an error depending on the direction of the magnetic field strength at the location of the observation, and have spectral resolutions that are worse than ground-based instrumentation. Testing using these datasets has provided good cross-calibration information for the magnetic fields measured by these instruments and BABO \citep{2014ApJ...787...22W}, and the purpose of this article is to take the calibrations to the next step.

The Facility InfraRed Spectropolarimeter (FIRS) instrument \citep{2010MmSAI..81..763J} installed at the Dunn Solar Telescope (DST) on Sacramento Peak, NM is a multi-slit imaging spectropolarimetric instrument with far higher spectral resolution than both MDI and HMI. By cross-calibrating data from FIRS and BABO, further testing of the BABO magnetic field measurements can be made and there are distinct advantages to using FIRS data as opposed to data from MDI or HMI. FIRS is able to operate at 1564.8~nm, the same wavelength as used in BABO observations. This means that the same reduction technique can be used and we do not need to calibrate for the effect of observing at a different wavelength.

The structure of this article is as follows. In Section~\ref{sect:data}, we outline the instruments used and give details on the data used in this study. Section~\ref{sect:analysis} shows how we analyse the data and we give the results of our comparisons between different aspects of the data for the purposes of cross-calibration in Section~\ref{sect:results}. We then finish by giving our conclusions in Section~\ref{sect:conclusions}.

\section{Data}\label{sect:data}
The two sets of data used in this article come from the BABO and FIRS instruments.

\subsection{BABO at the McMath-Pierce Solar Telescope Facility}
BABO is a single element, nitrogen cooled, InSb photovoltaic detector \citep{Hall:75} that is regularly used at the McMath-Pierce Solar Telescope facility. Its name comes from the nearby Baboquivari mountain, a prominent peak visible from Kitt Peak where the instrument is located. The detector is sensitive from 1000~nm to 5100~nm and so a filter is used to restrict the bandwidth such that the detector is not subjected to more light than it can handle. We observe the Fe {\sc i} line at 1564.8 nm in first order and the corresponding intensity is very high. The observation is made by finding the darkest part of a sunspot umbra and measuring the spectrum in that location \citep{2006ApJ...649L..45P}. The resulting spectrum shows the line split into three components as a result of the Zeeman effect. The Lande $g$-factor of this atomic transition ($g_{\rm L}=3$) is large enough that magnetic field strengths greater than around 1100~gauss (G) can be measured. There is no image correction system used and the data are taken under natural seeing conditions with an aperture of 2.5 $\times$ 2.5 arcsec.

\subsection{FIRS at the Dunn Solar Telescope}\label{sect:FIRS}
FIRS at the Dunn Solar Telescope is developed by the Institute for Astronomy at the University of Hawai'i and the National Solar Observatory. It can be operated in single or multi-slit mode at wavelengths of 630.2~nm and 1564.8~nm, both transition lines of Fe {\sc i}, as well as the He {\sc i} 1083.0~nm spectral line. We use it at 1564.8~nm to scan sunspots and the observations provide us with spatial maps of the sunspot in Stokes $I, Q, U$ and $V$. These data are corrected by adaptive optics at a sampling of around 0.3 arcsec \citep{rimmele2004}. The spectral sampling is 5.5~pm per pixel over a wavelength range of 1561.6 -- 1567.2~nm.

\subsection{Dates}
To compare the data from both BABO and FIRS, we are limited to when both instruments were observing the same target at around the same time. BABO observations are taken between five and seven days per month. The relevant data from FIRS were obtained as part of the National Solar Observatory's service mode operations at the Dunn Solar Telescope\footnote{\url{http://nsosp.nso.edu/dst/smex}}. This gave five days when concurrent data was available and is detailed in Table~\ref{table:data}.

\begin{table}
\caption{Concurrent data available from both BABO(McMath-Pierce) and FIRS(DST)}
\begin{tabular}{  c | c | c }
  Date & Spots observed & NOAA number \\ \hline
  23 January 2013 & 3 & 11660\\
  4 February 2013 & 1 & 11667\\
  17 October 2013 & 6 & 11861\\
  18 October 2013 & 5 & 11874\\
  31 October 2013 & 7 & 11884\\
\end{tabular}
\label{table:data}
\end{table}

\section{Data Analysis}\label{sect:analysis}
Two methods are used to obtain magnetic field strength values from the FIRS data. The first is to use the Zeeman effect described earlier with the observed Stokes $V$ spectral profile.

\begin{figure}
\centering
\includegraphics[width=0.7\textwidth]{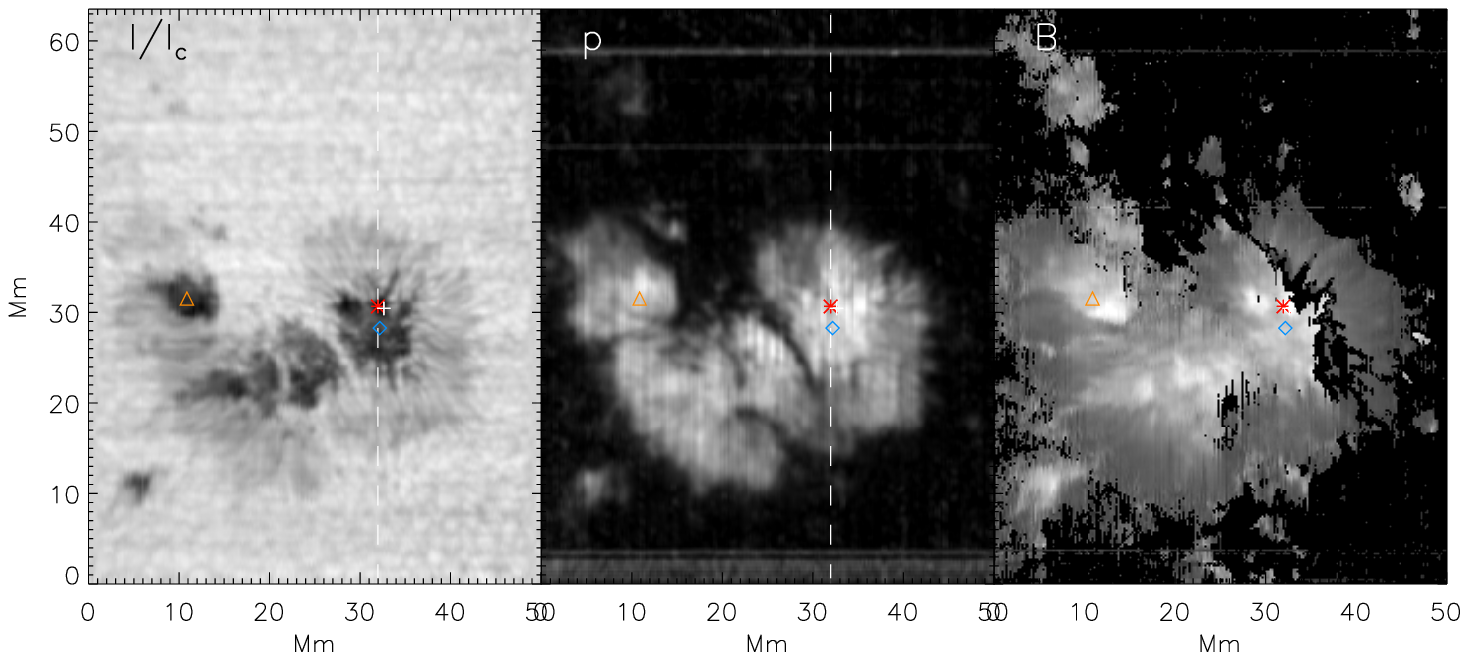}\includegraphics[width=0.18\textwidth]{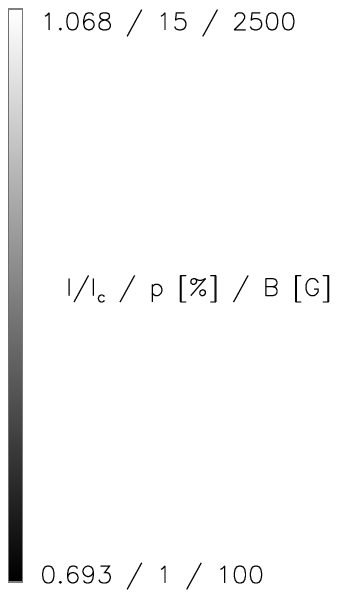}$ $ \\$ $ \\
\caption{Left to right: continuum intensity, polarization degree, and magnetic field strength. The vertical dashed lines denotes the location of the cuts shown in Figure~\ref{fig:FIRScuts}. The symbols denote the centre of a small umbral core (orange triangle), the centre of the biggest sunspot (blue diamond), the location of minimal intensity (red asterisk) and of maximal field strength (white cross). The intensities of the three panels are, from left to right, solar intensity, degree of polarisation, and magnetic field strength. The scales are shown by the bar on the right of the panels.}
\label{fig:FIRSfig}
\end{figure}

\begin{figure}
\includegraphics[width=0.9\textwidth]{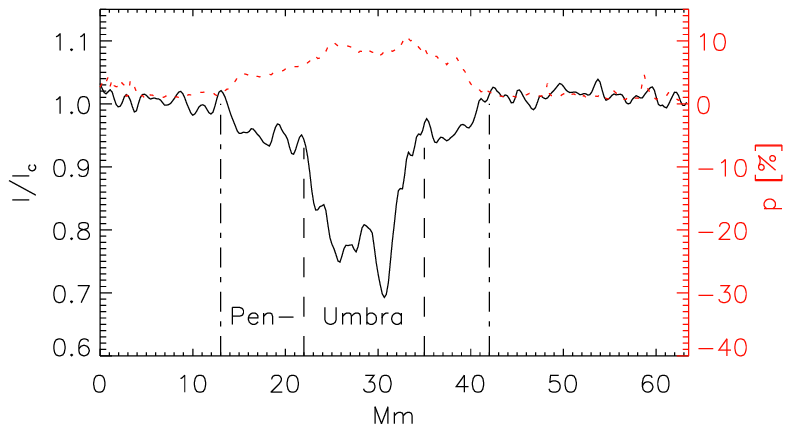}
\caption{Cuts across the sunspot along the dashed line of Figure~\ref{fig:FIRSfig}. Black (solid) and red (dotted) lines show: continuum intensity and polarization degree. The dash-dotted and dashed vertical lines denote penumbra and umbra, respectively.}
\label{fig:FIRScuts}
\end{figure}

The reason for using Stokes $V$ as opposed to Stokes $I$ (used by BABO) is due to the effects of stray light. When looking at a sunspot in Stokes $I$, the sunspot umbra is the darkest part of the image with a lighter penumbra surrounding it, and an even lighter quiet Sun around that. As such, the measurement of the umbra is affected by stray light from the brighter penumbra and quiet Sun regions. However, when observing the polarisation signal, the sunspot umbra is brighter than the penumbra and quiet Sun (cf.~Figures \ref{fig:FIRSfig} and \ref{fig:FIRScuts}). This means that the contributions from polarised stray light are significantly reduced when compared with the same measurement in Stokes $I$. 

To test the effects of the coarser spatial sampling or the reduced spatial resolution of BABO relative to FIRS, we degraded a FIRS data set from its original sampling of $0.3''$ to $2.5''$ (BABO aperture) and $5''$ (BABO aperture under bad seeing). We chose four different spatial locations. Three were located inside the biggest sunspot of the group at the pixel with the highest magnetic field strength, the lowest intensity, and at the centre of the sunspot (cf.~Figure~\ref{fig:FIRSfig}). A fourth position was set to the centre of one of the smaller umbral cores that were also measured with BABO. The locations of highest field strength and lowest intensity in the big sunspot are in close proximity to each other, but at quite some distance from the centre of the sunspot in one of the three small, dark umbral cores at the upper border of the umbra (Figure~\ref{fig:FIRSfig}). Similarly, the centre of the smaller sunspot does not contain the highest field strength. 

\begin{figure}
\centering
\includegraphics[width=0.9\textwidth]{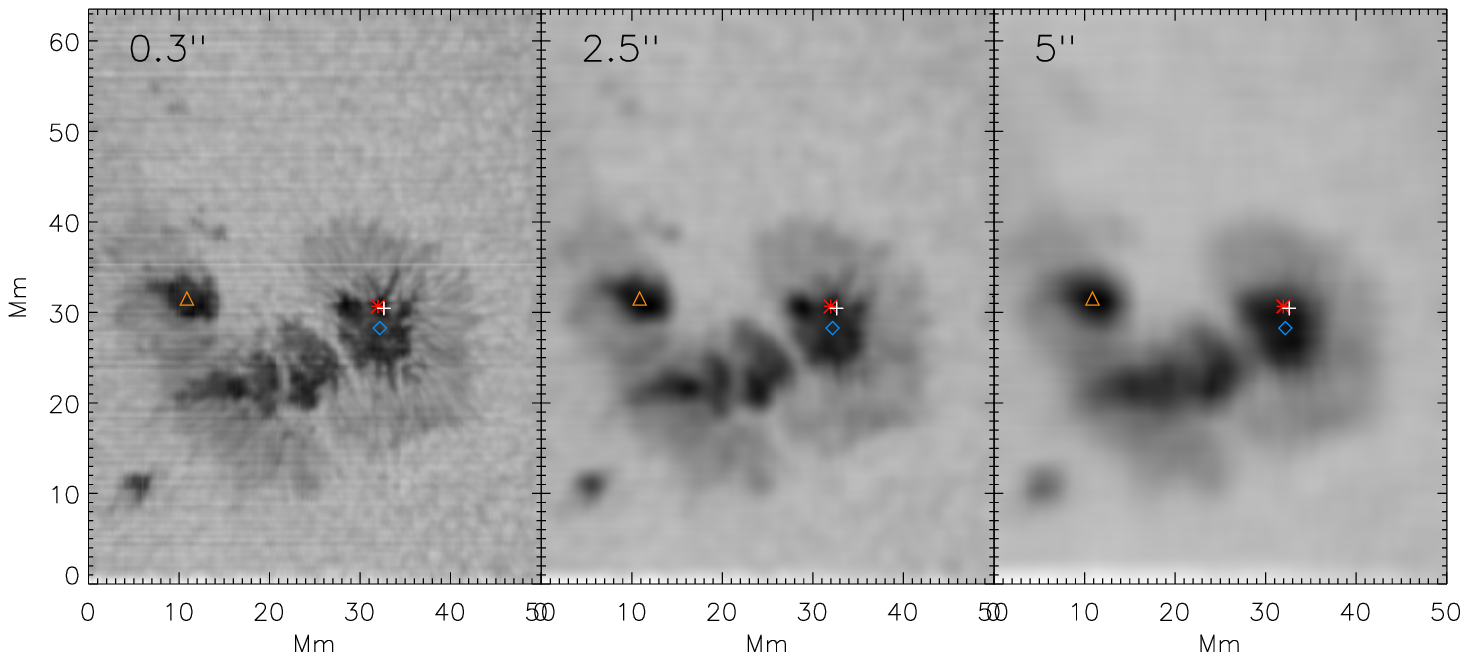}$ $\\$ $\\
\caption{Re-sampling of the observations. The left panel shows the original sampling while the middle and right panels are images re-sampled with 2.5$^{\prime\prime}$ and 5$^{\prime\prime}$ square pixels.}
\label{fig:resampling}
\end{figure}

\begin{figure}
\centering
\includegraphics[width=0.9\textwidth]{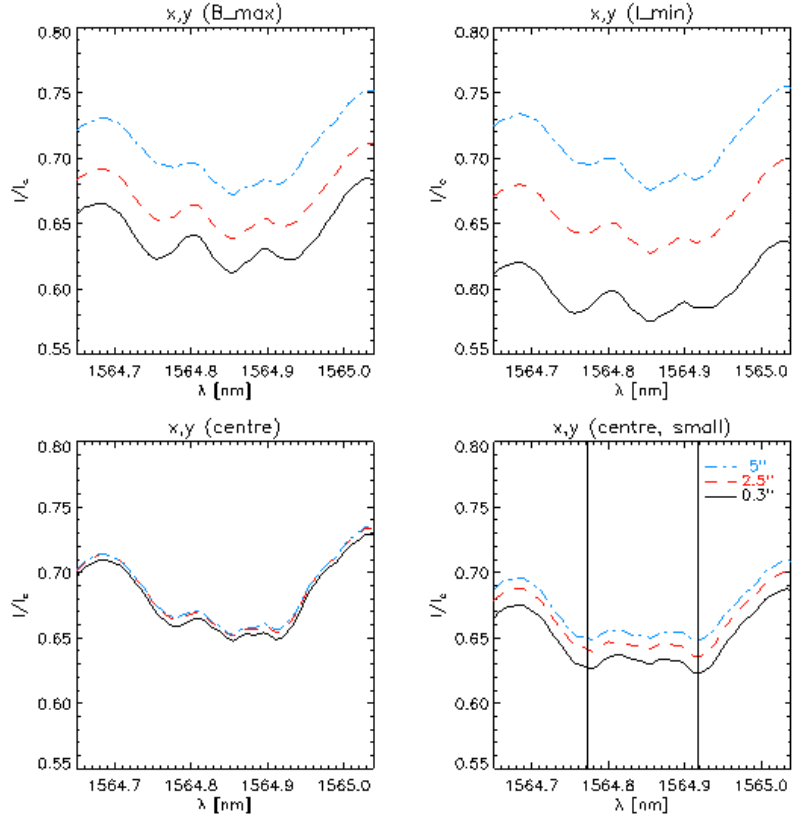}
\caption{Stokes $I$ profiles at four locations at different spatial sampling at maximal field strength (top left), minimal intensity(top right), centre of the biggest sunspot (bottom left), and centre of a small umbral core (bottom right). Black (solid), red (dashed), and blue (dotted) lines indicate sampling sizes of $0.3''$, $2.5''$, and $5''$, respectively. The black vertical lines indicate the location of the $\sigma$-components.}
\label{fig:stokesI}
\end{figure}

\begin{figure}
 \centering
\includegraphics[width=0.9\textwidth]{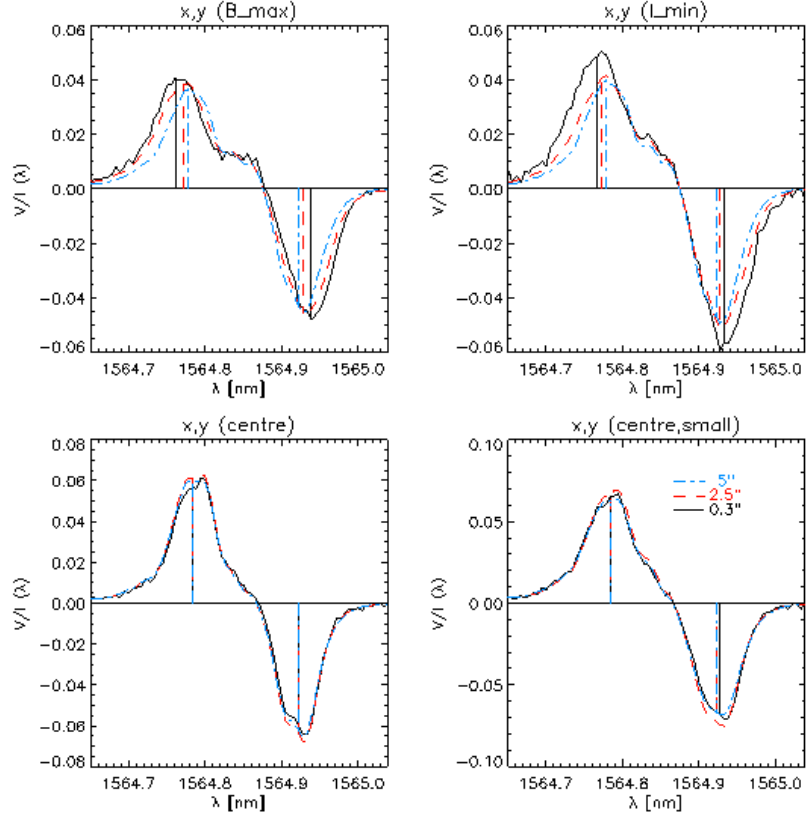}
\caption{Same as Figure~\ref{fig:stokesI} but with Stokes $V$ profiles instead of Stokes $I$ profiles. The black and coloured vertical lines indicate the location of the $\sigma$-components.}
\label{fig:stokesV}
\end{figure}

When the FIRS data are degraded, the location of lowest intensity in the large sunspot moves towards its centre (Figure~\ref{fig:resampling}). The smaller and darker umbral cores can no longer be distinguished at $5''$ sampling. Figures.~\ref{fig:stokesI} and \ref{fig:stokesV} show the Stokes $I$ and $V$ profiles at the four selected locations under different sampling. For both the pixels with highest field strength and lowest intensity in the large sunspot the splitting reduces visibly with the increasing degradation because of their proximity to the umbral boundary (top row of Figures~\ref{fig:stokesI} and \ref{fig:stokesV}). The profiles at the centre of the large and small sunspot change only slightly (bottom row of Figures~\ref{fig:stokesI} and \ref{fig:stokesV}). We then determined the magnetic field strength from the distance of the $\sigma$-components in the original and degraded Stokes $V$ profiles. Table \ref{tab:Bstrength} shows that the original values decrease by 300--400\,G for the first two locations, while the other two locations yield a nearly constant field strength regardless of the degradation. Note that, however, the field strength at the centre of the large sunspot is actually more than 500\,G lower than the maximum field strength inside the sunspot. The same applies for the smaller umbral core: the maximal field strength of this umbral core is located towards the west of its centre (Figure~\ref{fig:FIRSfig}) and reaches 2500\,G like in the bigger sunspot. 

\begin{table}
\caption{Magnetic field strength in G at four locations at different spatial sampling.}

\begin{tabular}{c|cccc}
Sampling &            $0.3"$  &    $2.5"$ &      $5"$     &     BABO\cr\hline
$x,y$ (max $B$) &         2528&      2335 &     2117    &      -  \cr     
$x,y$ (min $I$) &         2400 &     2226 &     2107    &      -  \cr
$x,y$ (centre) &        1987 &     1983 &     1982    &     1969\cr
$x,y$ (centre, small) & 2084 &     2024 &     2047    &     1931\cr
\label{tab:Bstrength}
\end{tabular}
\end{table}

The second method used for measuring magnetic fields strengths in FIRS data is by making use of spectropolarimetric inversion techniques. For this we use the Stokes Inversion based on Response functions (SIR) code developed by \citet{1992ApJ...398..375R}.

The SIR code was developed to handle the inversion of spectral lines formed in the presence of magnetic fields and can fit any combination of Stokes parameters for any number of given spectral lines. The method used involves taking the initial observations and a user-provided model atmosphere, which is then iterated until the synthetic Stokes profiles generated by the model match with the observed profiles. The result is the thermal, dynamic, and magnetic structure of the iterated model atmosphere and allows physical observations of solar parameters to be made. In our case, the parameter we are interested in is the magnetic field strength at the location where the spectral line is formed.

\subsection{Comparison of Magnetic Fields}\label{sect:compare}

\begin{figure}[ht!]
  \centering
  \includegraphics[width=0.9\textwidth]{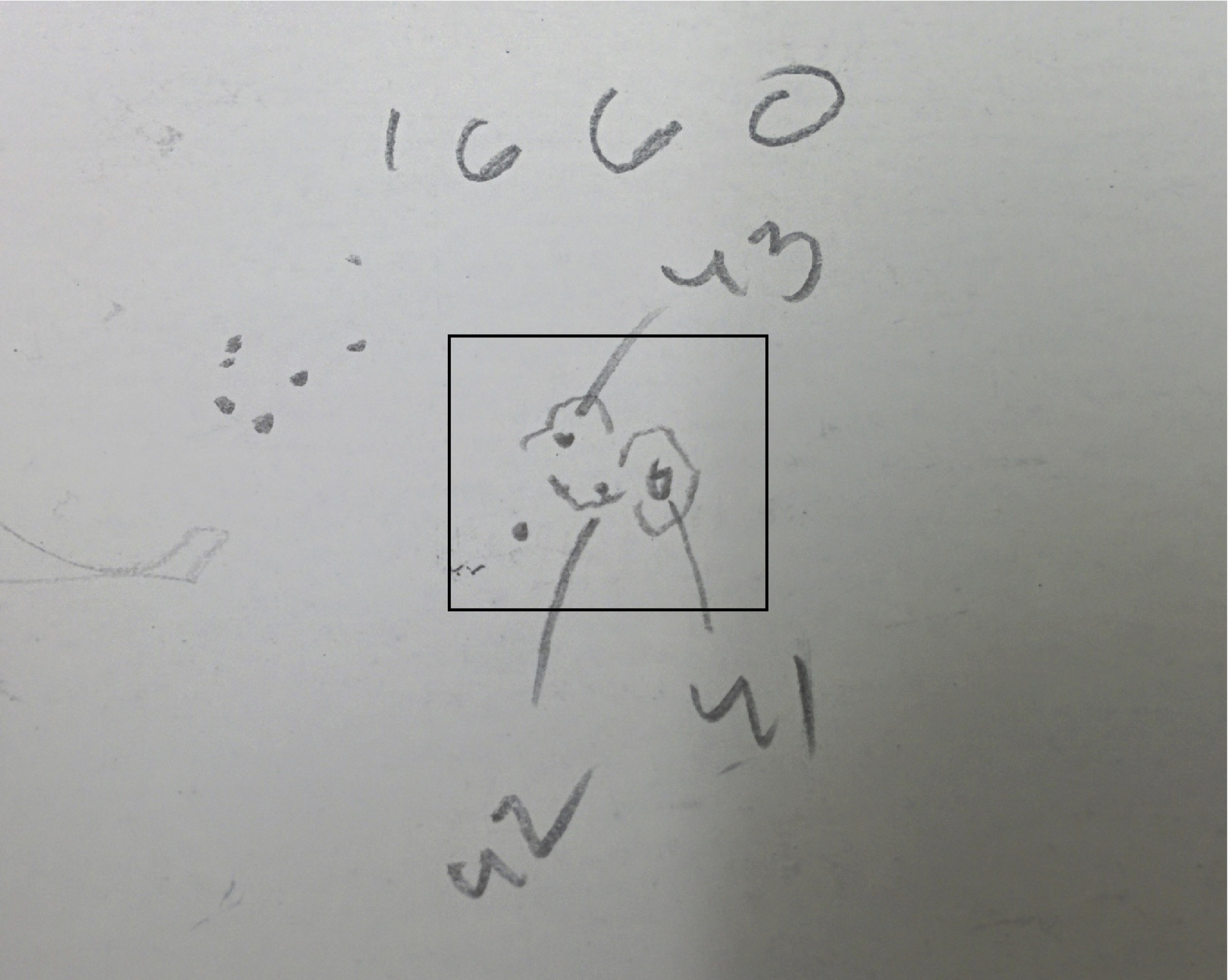}
  \caption{Drawing from the west auxilliary telescope at the McMath-Pierce Solar Telescope facility on 23 January 2013. The drawing shows NOAA active region 11660 and marks the location of three umbral measurements: u1, u2, and u3. The box in the centre shows the approximate area of the FIRS scan in Figure~\ref{fig:FIRSfig}.}
  \label{fig:BABO_drawing}
 \end{figure}

When comparing magnetic fields from the two datasets, it is crucial that the locations of the observations are known and understood. For the BABO data, this is relatively difficult. The BABO detector is a single pixel (a diode) and so the aperture size of the instrument setup is effectively the pixel size. This aperture subtends an angle of 2.5 by 2.5 arcsec on the solar disk, which is far larger than the angle subtended by FIRS pixels (and the angle changes depending on the observation and scanning speed of FIRS). As such, the FIRS spectra must be averaged over the target area to provide a fair comparison to the BABO data, as mentioned in Section~\ref{sect:FIRS}. Later in this section, we also show the comparison when the FIRS data are not averaged to show the difference. With BABO, the observation is taken at the darkest point in the sunspot umbra as measured at the telescope, but the exact location of the observation is not noted. A white-light drawing of the Sun is made before the observations are taken and the umbral spectra obtained are named based on the umbra measured. An example of one of these drawings is shown in Figure~\ref{fig:BABO_drawing} and is of the same region seen in the FIRS data in Figure~\ref{fig:FIRSfig}.

 By averaging the FIRS spectra before finding the darkest pixel in the umbra, we do not find the darkest pixel in the original FIRS data, but the darkest area subject to the aperture size. The darkest part of the umbra measured in this way is usually at the centre, farthest from the penumbra, even though it may not include the darkest FIRS pixels in Stokes $I$. This effect was demonstrated in Section~\ref{sect:FIRS}. This is because the aperture size is large enough that a few very dark pixels are not sufficient to affect the average intensity over the whole area. If the darkest pixels in the umbra are near the penumbra, the overall brighter intensity of that area means that, on average, the centre of the umbra is darker and so will be the location of the BABO measurement.

 \section{Results}\label{sect:results}
\begin{figure}[ht!]
  \centering
  \includegraphics[width=0.9\textwidth]{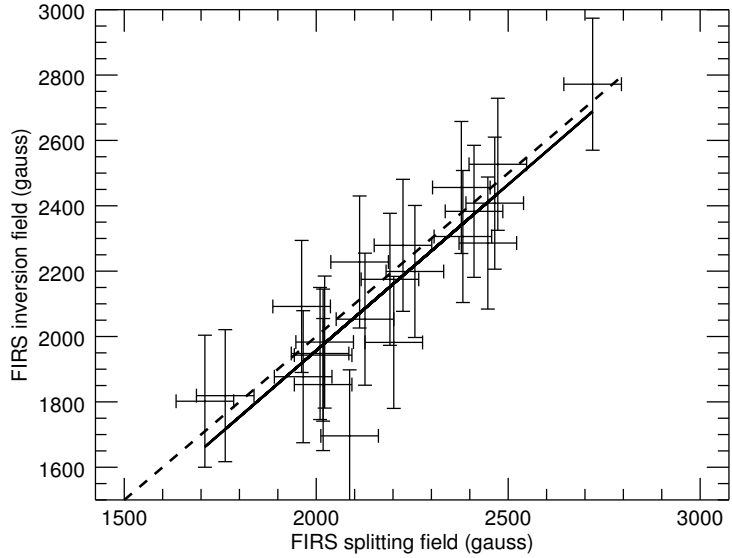}
  \caption{Magnetic field strengths obtained from FIRS data by inversion, and by Zeeman splitting. The solid line shows the best fit to the data and the dashed line is $y = x$.}
  \label{fig:FIRS_v_FIRS}
 \end{figure}

As mentioned in Section \ref{sect:FIRS}, two different methods were used to obtain magnetic field strength values from FIRS data. The relationship between magnetic field strengths calculated from Zeeman splitting, and from the SIR inversion code was tested and the results are shown in Figure~\ref{fig:FIRS_v_FIRS}. The plot shows that there is a relationship between the magnetic field strength values returned by the two methods, as we would expect. The best fitting line through the data is very close to the line $y = x$ and shows that, over this set of data, both methods are measuring the same magnetic field strengths. The linear fit parameters for all of the figures in this article are given in Table~\ref{table:fitparams} for reference.

\begin{figure}[ht!]
  \centering
  \includegraphics[width=0.9\textwidth]{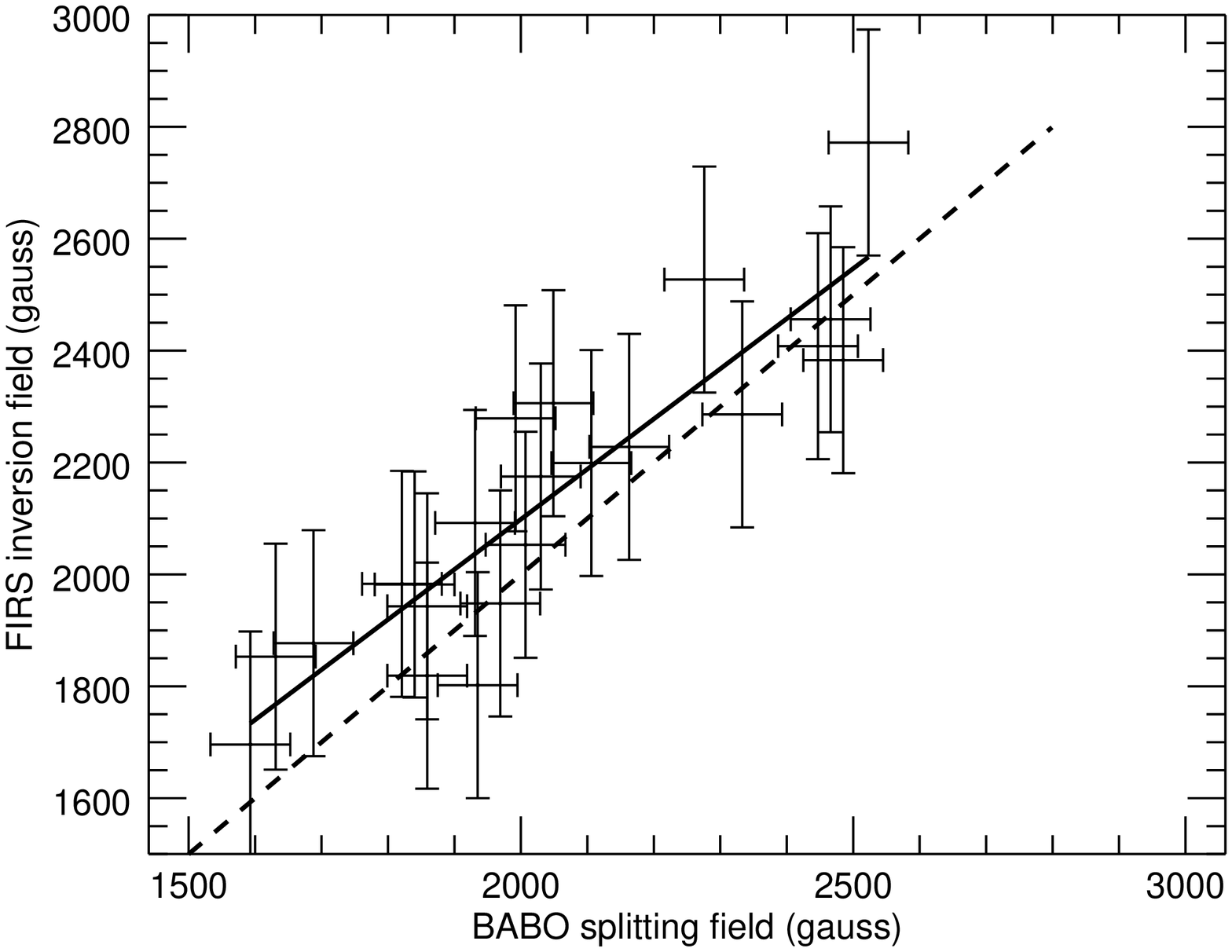}
  \caption{Magnetic field strengths obtained from Zeeman splitting of Stokes $I$ BABO data, and inversions of FIRS data. The solid line shows the best fit to the data and the dashed line is $y = x$.}
  \label{fig:FIRS_v_BABO}
 \end{figure}

When comparing the data from FIRS and BABO with one another, we only used the magnetic fields obtained by inversion from FIRS. The reason for this is that the information from all four Stokes parameters $IQUV$ is used by the inversion, not only Stokes $I$ or $V$. For sunspots far from disk centre, the neutral line of Stokes $V$ may pass through the umbra, which makes the splitting of $V$ unreliable (see Figure~\ref{fig:FIRSfig}). Figure~\ref{fig:FIRS_v_BABO} shows this comparison in detail.
 
The agreement between these data is still good, although not as good as the agreement between magnetic fields from FIRS data by two different methods.

\begin{figure}[ht!]
  \centering
  \includegraphics[width=0.9\textwidth]{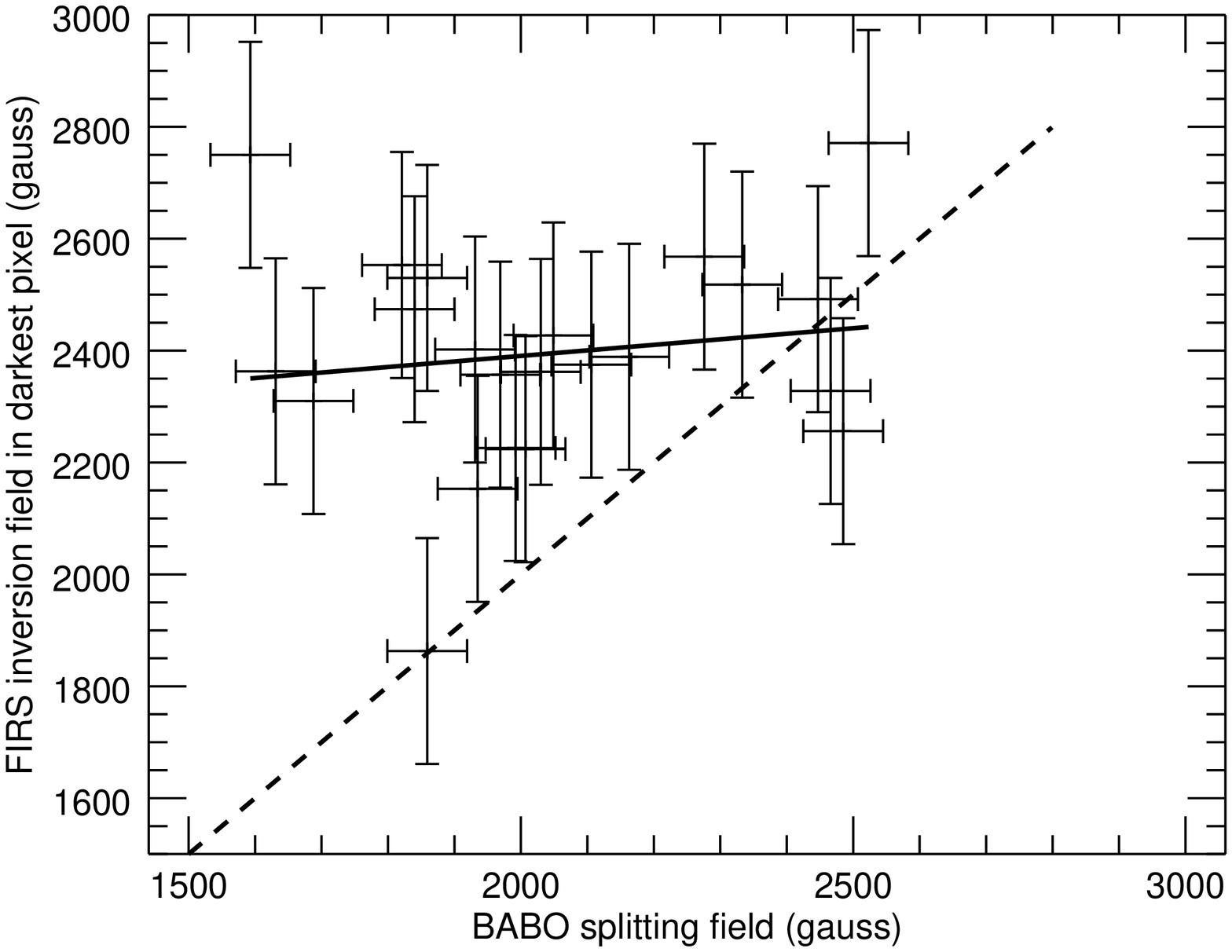}
  \caption{Magnetic field strengths obtained from Zeeman splitting of Stokes $I$ BABO data, and inversions of FIRS data in the location of the darkest umbral pixel. The solid line shows the best fit to the data and the dashed line is $y = x$.}
  \label{fig:FIRS_v_BABO_max}
 \end{figure}

It was mentioned that the FIRS data used so far do not give the true maximum umbral magnetic field due to the constraint of the BABO aperture size being applied. To show the effect of this, in Figure~\ref{fig:FIRS_v_BABO_max} we show the relationship between the BABO magnetic field strength and the FIRS magnetic field strength in the darkest location of the umbra in the FIRS Stokes $I$ data. This is how the location of BABO measurements is decided and is the fairest comparison, although the spatial resolution of the observations are very different.
 
We see that changing the spatial resolution of the FIRS data in this comparison has a huge effect on the relationship between the two data sets. There is no trend between the BABO and FIRS data, with the FIRS magnetic field value being similar regardless of the BABO magnetic field value. As such, we must take great care when comparing these two sources of data as the spatial resolution over which the magnetic field strength is evaluated is critical.

  \begin{figure}[ht!]
  \centering
  \includegraphics[width=0.9\textwidth]{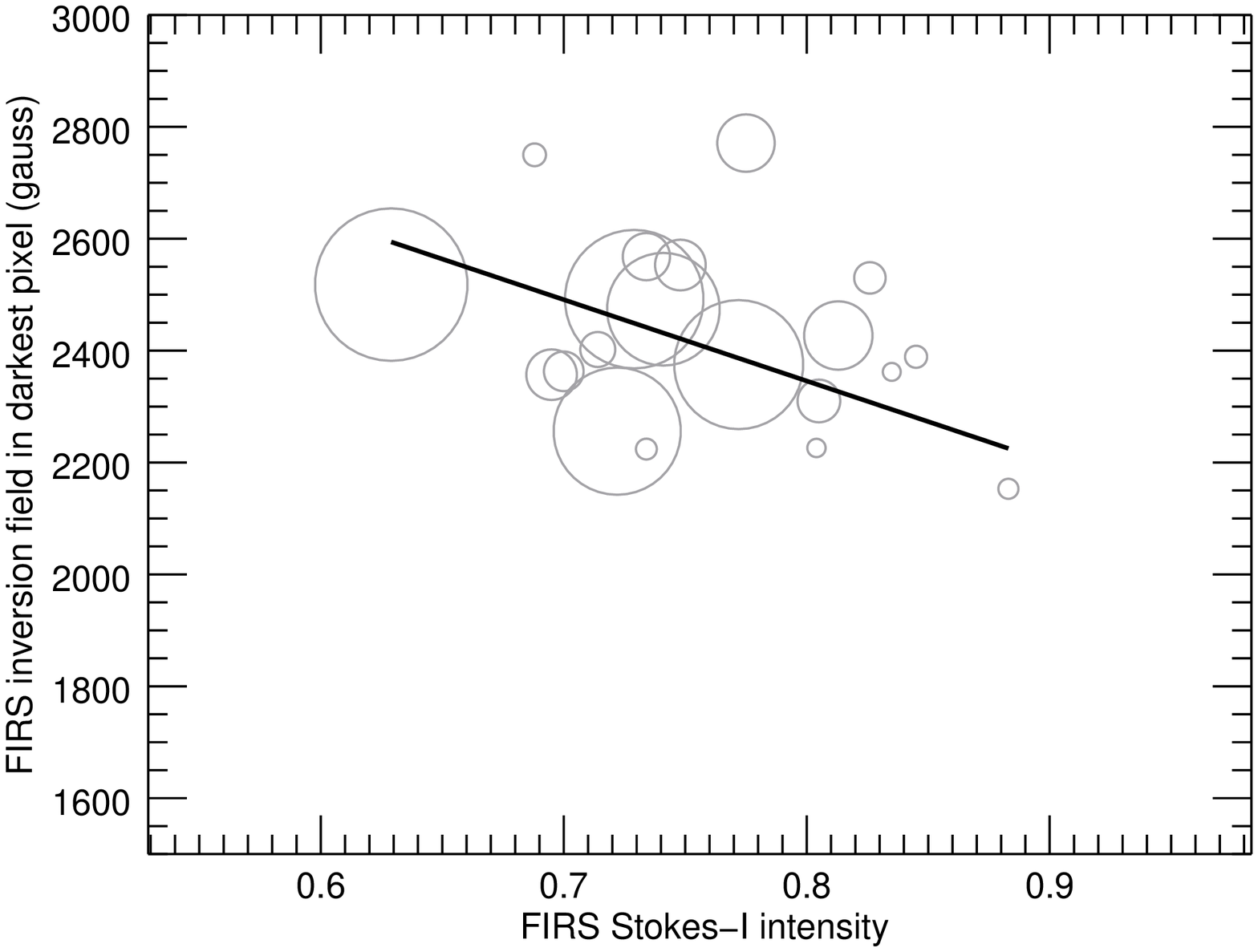}
  \caption{Intensities from the darkest umbral pixel in Stokes $I$ FIRS data, and the magnetic field strengths of the the same pixel from FIRS inversions. An intensity of 1.0 corresponds to the average intensity in a quiet Sun area. The area of the circles used are proportional to the areas of the observed sunspot umbrae.}
  \label{fig:B_v_I}
 \end{figure}
 
Finally, the relationship between the magnetic field strength in the darkest FIRS umbral pixel and the intensity of the darkest FIRS umbral pixel is examined. It has previously been shown that as the magnetic field strength increases, the intensity of solar plasma decreases (\citealp{1993A&A...270..494M}; \citealp{1994IAUS..154..477K}) and we previously showed strong correlation between these quantities for BABO, SOHO/MDI and SDO/HMI data in \citet{2014ApJ...787...22W}. The FIRS data for the sunspots observed in this study is shown in Figure~\ref{fig:B_v_I}. Although we see a weak correlation between these parameters (with $R^2 = 0.23$), the scatter is very large. Based on previous studies, it is likely that the poor correlation is due to the small sample size of sunspots used in this article.

.

\begin{table}
\caption{Parameters from linear fits.}
\begin{tabular}{  c | c  c  c }
  Figure & Intercept (error) & Gradient (error) & $R^2$ \\ \hline
  \ref{fig:FIRS_v_FIRS} & -74 (242) & 1.01 (0.11) & 0.80\cr
  \ref{fig:FIRS_v_BABO} & 305 (200) & 0.90 (0.10) & 0.81\cr
  \ref{fig:FIRS_v_BABO_max} & 2192 (323) & 0.10 (0.16) & 0.02\cr
  \ref{fig:B_v_I} & 3508 (453) & -1453 (590) & 0.2\cr
\end{tabular}
\label{table:fitparams}
\end{table}

\section{Conclusions}\label{sect:conclusions}
We have undertaken a cross-calibration using data from BABO at the McMath Pierce Solar Telescope and FIRS at the Dunn Solar Telescope. We obtained the magnetic field from FIRS data in two different ways, one being Zeeman splitting of the Stokes $V$ signal and the other being inversion of the Stokes parameters. It was found that both of these methods gave very similar values for the magnetic field strength.

Next, the field strengths from FIRS were compared to those from BABO, taking care to match up the effective pixel size due to the larger aperture and single element design of BABO. A strong correlation was found between the BABO and FIRS magnetic field strengths, providing a reliable calibration between the two datasets and allowing for FIRS data to also gather observations for the synoptic catalogue.

We then examined the effects of the pixel size of the observations as the effective BABO pixel size is far larger than that of FIRS. If this difference is not taken into account, there is no correlation between the observations from both instruments.

Finally, the FIRS umbral magnetic field strength and intensity were compared to look for the a previously well established relationship, but a weak correlation was found. We propose that this is due to the small number of sunspots that have been used in this study.

These results show that, as long as care is taken, the umbral magnetic field measurements from FIRS are well correlated with those from BABO. The synoptic measurements could thus be continued with FIRS observations of sunspots in the future.

\section*{Acknowledgments}
We thank the referee for their constructive comments which have served to improve the quality of this manuscript.  The McMath-Pierce Solar Telescope facility and the Dunn Solar Telescope are operated by the National Solar Observatory. The National Solar Observatory is operated by the Association of Universities for Research in Astronomy under a cooperative agreement with the National Science Foundation. This publication makes use of data obtained during Cycles 1 and 2 of the DST Service Mode Operations under the proposal IDs P107 and P509.
 
\bibliographystyle{spr-mp-sola}
\bibliography{bibliography}

\end{document}